\documentclass[prl,,nofootinbib,twocolumn,superscriptaddress]{revtex4}
\usepackage{epsfig}
\usepackage{axodraw}

\newcommand{\la}{{\lambda}}

\newcommand{\be}{\begin{equation}}
\newcommand{\ee}{\end{equation}}
\newcommand{\beq}{\begin{equation}}
\newcommand{\eeq}{\end{equation}}
\newcommand{\bea}{\begin{eqnarray}}
\newcommand{\eea}{\end{eqnarray}}
\newcommand{\br}{\begin{eqnarray}}
\newcommand{\er}{\end{eqnarray}}
\newcommand{\ba}{\begin{array}}
\newcommand{\ea}{\end{array}}
\newcommand{\bi}{\begin{itemize}}
\newcommand{\ei}{\end{itemize}}
\newcommand{\bn}{\begin{enumerate}}
\newcommand{\en}{\end{enumerate}}
\newcommand{\bc}{\begin{center}}
\newcommand{\ec}{\end{center}}

\newcommand{\bT}{\bar{T}}
\def\bY{{\bf Y}}
\def\bA{{\bf A}}
\def\bB{{\bf B}}
\def\bP{{\bf P}}
\def\bU{{\bf U}}
\def\bV{{\bf V}}
\def\bm{{\bf m}}
\def\bM{{\bf M}}

\def\tl{{\tilde{L}}}
\def\tm{{\tilde{m}}}

\def\tq{{\tilde{Q}}}

\def\unity{{\hbox{1\kern-.8mm l}}}

\newcommand{\no}{\nonumber}

\newcommand{\ga}{\gamma}

\newcommand{\gsim}{\lower.7ex\hbox{$\;\stackrel{\textstyle>}{\sim}\;$}}
\newcommand{\lsim}{\lower.7ex\hbox{$\;\stackrel{\textstyle<}{\sim}\;$}}

\def\mysection#1{\noindent {\bf #1} }

\begin{document}


\title{\bf
A Predictive Seesaw Scenario for EDMs}

\author{Eung Jin Chun}
\affiliation{Korea Institute for Advanced Study, Seoul 130-722,
Korea}
\author{Antonio Masiero}
\author{Anna Rossi}
\affiliation{Dipartimento di Fisica ``G.~Galilei'', Universit\`a di Padova
and INFN, Sezione di Padova, I-35131 Padua, Italy}
\author{Sudhir K. Vempati}
\affiliation{Dipartimento di Fisica ``G.~Galilei'', Universit\`a di Padova
and INFN, Sezione di Padova, I-35131 Padua, Italy}
\affiliation{Centre de Physique Theorique, Ecole Polytechnique-CPHT,
91128 Palaiseau Cedex, France
}

\date{\today}

\vspace*{0.in}

\begin{abstract}
The generation of 
electric dipole moments (EDMs) is addressed in the supersymmetric seesaw 
scenario realized through  the exchange of $SU(2)_W$ triplet states. 
In particular, we show that 
the triplet soft-breaking  bilinear term can induce 
finite  contributions to 
lepton and quark EDMs. Moreover, 
the peculiar flavour structure of the model allows us to predict the 
EDM ratios $d_e/d_\mu$ and $d_\mu/d_\tau$ only in terms of 
the neutrino parameters.   

\end{abstract}

\maketitle

Supersymmetric (SUSY) extensions of the Standard Model  exhibit
plenty of new CP violating phases in addition to the unique CKM phase 
of the SM. Although new sources of CP violation are welcome to 
dynamically achieve an adequate matter -- antimatter asymmetry, it is 
known that they constitute a threat for very sensitive CP tests like
those of the electric dipole moments (EDMs), at least for SUSY masses 
which are within the TeV region. In view of such constraints as well as 
of those coming from flavour changing neutral current  processes, one 
can safely assume that all the terms which softly break SUSY are real and 
flavor universal at the scale where supersymmetry breaking is communicated
to the visible sector. In spite of this, their renormalization group (RG)  
running down to the electroweak scale feels the presence of the Yukawa 
couplings in the superpotential which can induce flavor and CP violation in 
the soft breaking sector at low energy. Hence,    
in this class of  minimal supersymmetric extensions of the Standard Model 
(MSSM), the CP and flavour  violating radiative effects are effectively 
encoded in the CKM mixing matrix, and therefore can be within 
the experimental constraints. 

This picture can drastically  change if there are additional 
sources of flavour violation in the superpotential. This is what 
happens in the case of the  seesaw mechanism which entails    
new lepton flavour violating (LFV) and CP violating (CPV)  
Yukawa couplings to generate neutrino masses  \cite{lfv}.
Regarding the standard (so-called type-I) seesaw mechanism \cite{seesaw}, 
the phenomenological implications of the RG-induced  LFV and CPV
effects for 
rare decays such as $\mu \to e + \gamma$ 
and  leptonic EDMs  have been studied in detail 
in several works \cite{rgess,peskin}. 

In addition to the logarithmically divergent RG corrections,    
there are additional finite contributions to the soft SUSY  terms 
in the seesaw mechanism.  In type-I seesaw, 
these contributions are induced by 
the bilinear soft-term $\bB_N \bM_N \tilde{N}\tilde{N}$, associated
with the Majorana mass matrix $\bM_N$  for the heavy singlet states $N$. 
The fact that they can be significant 
and even comparable to the usual RG-induced contributions has been recently 
shown in Ref. \cite{farzan}. If the only source of
CPV resides in the soft matrix $\bB_N$, then 
 the most stringent constraint  arises from the
lepton EDMs.
We recall that the flavour structure of  the  infinite 
radiative 
contributions to the soft-breaking terms  is determined by the Dirac-like 
Yukawa couplings 
$\bY_N$ of the $SU(2)_W$ lepton doublets with the heavy states $N$.  
In general  the  
Yukawa couplings $\bY_N$ are arbitrary complex parameters and, moreover, 
not directly related to the low-energy neutrino mass matrix $\bm_\nu$. 
Regarding  the finite radiative 
corrections, they  exhibit in general  
a different flavour structure from the infinite one, 
as the soft parameter 
$\bB_N$ is a symmetric matrix in the generation space, not related 
to the neutrino mass matrix $\bm_\nu$. 
Hence both the above radiative corrections cannot be 
predicted in terms of known low-energy observables \cite{DI}.

In this Letter we propose a predictive alternative  
by considering the triplet 
seesaw mechanism where neutrino masses are generated through the exchange 
of one pair of $SU(2)_W$  triplet states $T, \bar{T}$ 
with nonzero hypercharge  (this realization  
is similar to what was initially 
suggested in Refs. \cite{ss2,ma} and, for concreteness, 
we consider here the supersymmetrized version in Ref. \cite{ar})
Indeed, in this case the flavour structure of the (finite and infinite) 
radiative 
corrections can be rewritten in terms of the 
low-energy neutrino masses and mixing angles up to an overall mass scale. 
Furthermore, assuming no new SUSY CP violating phases, leptonic EDMs 
can be shown to be proportional to the smallest neutrino mixing angle,
$\theta_{13}$. 
Here we propose a supersymmetry
breaking scenario where the soft-term $B_T M_T {T} \bar{T}$ is 
the {\it only} new source of CP violation.
In such a situation large leptonic and hadronic EDMs can be generated, 
for a reasonable SUSY mass spectrum, and  
interestingly enough,  the ratios of the resultant leptonic 
EDMs are completely determined by the low-energy neutrino data and are 
independent of the soft spectrum. Such a  predictive power is 
directly linked to the above assumption on the uniqueness of the CPV 
phase. In view of our ignorance on the SUSY breaking mechanism, 
we are led to make assumptions in order to reduce 
the number of CP phases in a generic MSSM frame. 
This is what we do here with the advantage that our assumption 
leads to testable predictions as we will show. 

In the  type-I seesaw mechanism the superpotential reads:
\be
W =W_0 + W_N 
\label{seesawn}
\ee 
with 
\bea
&&W_0 = \bY_e H_1 e^c L
+\bY_d H_1 d^c Q + \bY_u H_2 u^c Q  + \mu H_1 H_2 \no \\
&& W_N = \bY_N H_2 N L + \frac12 \bM_N NN , \label{mssm}
\eea
where we have used the standard notation and  
family indices are understood. 
 Assuming flavor universality and CP conservation 
 of the SUSY sector,   
 finite \cite{farzan} and infinite \cite{peskin} LFV and CPV  radiative 
corrections are induced by the 
new flavour structures 
($\bY_N$ and $\bM_N$). Such contributions  are proportional to 
the quantity $\bY^{\dagger}_N \bY_N$. In spite of this dependence on the 
``leptonic'' quantity $\bY_N$, it is  worthwhile emphasizing that  these
contributions affect also the hadronic EDMs, a point 
which was missed in the literature. In particular, 
this applies to the finite contributions  to the trilinear
 $\bA_u$ term which is corrected as:
\be \label{au-n}
\delta \bA_u = - \frac{1}{16\pi^2} \bY_u
{\rm tr} (\bY^\dagger_N \bB_N \bY_N),
\ee
leading to quark EDMs, 
and thus to a nonzero neutron EDM.
The scheme yields the following ratios of EDMs: 
\be \label{edmI}
 {d_\mu \over d_e} \approx {m_\mu \over m_e} { (\bY^\dagger_N \bB_N \bY_N)_{22}
  \over (\bY^\dagger_N\bB_N \bY_N)_{11} } ,\quad
  {d_u \over d_e} \approx  C {m_u \over m_e} { {\rm tr}(\bY^\dagger_N \bB_N 
\bY_N)
  \over (\bY^\dagger_N  \bB_N \bY_N)_{11} },
\ee
where $C$ is a factor depending on the soft mass parameters.
The above ratios (\ref{edmI}) are strongly 
model-dependent given their dependence on the combination 
$\bY^\dagger_N \bB_N \bY_N$ \footnote{
The RG evolution spoils the possible (high-scale) flavour blind structure 
of the matrix 
$\bB_N$ through the contributions of the 'flavoured' trilinear couplings 
$\bA_N H_2 \tilde{N} \tl$.}.

A more predictive picture for LFV and CPV can emerge in the triplet 
seesaw case \cite{ma,ar}.
Here  the MSSM superpotential $W_{0}$ is augmented by:
\be
W_T\!=\!\!
\frac{1}{\sqrt{2}}(\bY_T L  T L +
\lambda_1 H_1 T H_1 + \lambda_2 H_2\bar T H_2)+ M_T T \bar T
\label{Wt}
\ee
where the supermultiplets $T\!\!=\!\!(T^0, T^{+},T^{++})$,
$\bar T\!=\!(\bT^{0}, \bT^{-}, \bT^{--})$ are in a vector-like
$SU(2)_W \times U(1)_Y$ representation, $T\sim (3,1)$ and $\bT \sim(3,-1)$.
$\bY_T$, a complex symmetric matrix, is characterized by 
   6  independent moduli  and 3 physical phases, while
the parameters $\la_2$
and $M_T$ can be taken to be  real, and  $\la_1$  is in general complex.
After  integrating out the triplet states at the scale $M_T$,
the resulting neutrino
mass matrix becomes
\be
\label{numass}
 \bm_\nu =\bU^* \bm^D_\nu \bU^\dagger = 
 \frac{v^2_2 \la_2}{M_T}\, \bY_T  ,
\ee
\begin{figure}[t]
\begin{center}
\begin{picture}(150,60)(-75,-30)
\DashArrowLine(-65,20)(-34,0){3}
\DashArrowLine(-65,-20)(-34,0){3}
\DashArrowArcn(0,-20)(40,150,30){3}
\DashArrowArc(0,20)(40,210,270){3}
\DashArrowArcn(0,20)(40,330,270){3}
\DashArrowLine(65,0)(34,0){3}
\Text(-75,20)[]{$\tilde{e}^c$}
\Text(-75,-20)[]{$H_1$}
\Text(-37,18)[]{$Y_e Y^{\dagger}_T$}
\Text(0,12)[]{$\tl$}
\Text(-15,-7)[]{$T$}
\Text(15,-7)[]{$\overline{T}$}
\Text(0,-20)[]{$\times$}
\Text(-34,0)[]{$\bullet$}
\Text(34,0)[]{$\bullet$}
\Text(3,-30)[]{$B_T M_T$}
\Text(46,12)[]{$Y_T M_T$}
\Text(75,0)[]{$\tl$}
\end{picture}
\end{center}
%
\caption{\em Example of one-loop finite contribution to the trilinear 
coupling $\bA_e H_1 \tilde{e}^c \tl$, induced by the bilinear term
$B_T M_T T \bT$.}
\end{figure}
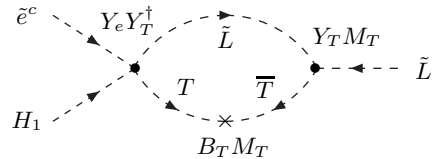
where $\bm^D_\nu$ is the diagonal neutrino mass matrix and $\bU$ is the PMNS 
leptonic mixing matrix, which is parametrized 
as $\bU = \bV[\theta_i,\delta] 
\bP[\phi_a]$ ($\theta_i, i=1,2,3$ are the mixing angles, $\delta$ is the 
Dirac phase, while $\phi_a, a=1,2$ are the Majorana phases). 
Eq.~(\ref{numass}) shows that 
the nine independent parameters contained in $\bY_T$
are  in {\it one-to-one} correspondence with the low-energy neutrino
parameters described by the quantities $\bU$ and $\bm^D_\nu$. 
As a consequence unambigous predictions
on the low-energy LFV phenomena can be derived 
in the triplet seesaw model \cite{ar,chun}.
We now turn to the EDM predictions in this model. First of 
all, out of the three phases present in the neutrino sector, only the Dirac 
phase $\delta$ may entail CP-violating effects
in the LFV entries (this is due to the symmetric nature of $\bY_T$).
However, the contributions to physical observables such as the EDMs
turn out to be quite suppressed in general.  
Indeed, due to the hermeticity of  $\bY^\dagger_T \bY_T$,
the phase of the electron EDM amplitude is always proportional to the small 
neutrino mixing angle $\theta_{13}$ and to a high power of the Yukawa 
couplings. Only in very special circumstances with $\theta_{13}$ close to 
the present experimental limit and very large $\tan\beta$, these 
contributions could become sizeable.

On the other hand,  a single CP phase residing in the soft term  $B_T M_T 
T\bar{T}$
can play a significant role in generating non-zero EDMs, 
once  we assume  vanishing CP phases in $\mu$ and
tree-level $A$-terms.
In such a case,
the trilinear couplings $\bA_e, \bA_d, \bA_u$ receive finite `complex'
radiative corrections at the decoupling of the heavy states $T, \bar{T}$,
exhibiting the common phase from the soft-term $B_T$\footnote{  
In fact, the $B_T$-term also induces finite
complex corrections to the Higgs bilinear term $B \mu H_1 H_2$ 
proportional to $(|\lambda_1|^2 + |\lambda_2|^2)$. However, the 
effect of the related CPV phase on the leptonic EDMs can be 
suppressed if {\it e.g.} the 
combination   $(|\lambda_1|^2 + |\lambda_2|^2)$ is much smaller 
than $\bY^\dagger_T \bY_T$. 
In this case the dominant contributions 
to the leptonic 
EDMs are those driven by the trilinear couplings as discussed here. 
}.
In Fig.~1 we show the  diagrammatic contribution to  $\bA_e$ proportional to 
$\bY^\dagger_T \bY_T$. Similar diagrams generate other contributions 
proportional to $|\lambda_i|^2$, 
relevant for  $\bA_e, \bA_d$ and  $\bA_u$.
Thus we obtain: 
\bea
 \delta {\bA}_e & = &
-\frac{3}{16 \pi^2} \bY_e \left(\bY^\dagger_T \bY_T
+ |\la_1|^2\right) B_T , \nonumber\\
\delta {\bA}_d & = &
-\frac{3}{16 \pi^2} \bY_d  |\la_1|^2 B_T,\label{finite} \\
\delta {\bA}_u & = &
-\frac{3}{16 \pi^2} \bY_u  |\la_2|^2 B_T . \nonumber
\eea
The  lepton (quark) EDMs arise from  one-loop diagrams that involve
the exchange of  sleptons (squark) of both chiralities  and Bino (gluino)
(at leading order in  the electroweak breaking effects).
The parametric dependence of the EDMs
at the leading order in the trilinear couplings
goes as follows 
\bea
\label{edmsall}
\frac{(d_{e})_i}{e }& \approx&  \frac{-\alpha}{4\pi c^2_W} m_{e_i} 
\frac{ M_1{\rm Im} (\delta \hat{ \bA}_{e})_{ii} }{m^4_\tl} F(x_1) 
   ,\nonumber\\
\frac{(d_{d})_i}{e}& \approx &  \frac{-2\alpha_s}{9 \pi} m_{d_i} 
\frac{M_3 {\rm Im} (\delta \hat{\bA}_d)_{ii}}{m^4_\tq} F(x_3) , \label{edm}\\
\frac{(d_{u})_i}{e}& \approx &  \frac{4 \alpha_s}{9 \pi} m_{u_i} 
\frac{M_3{\rm Im} (\delta \hat{\bA}_u)_{ii}}{m^4_\tq} F(x_3) ,  \nonumber
\eea
where $M_1$ and $M_3$ are the Bino and gluino masses, respectively, 
the trilinear couplings have been parametrized as
$\delta {{\bA}}_f = \bY_f \delta \hat{\bA}_f (f= e, u , d)$, and   
 $F(x)$  ($x_1= {M_1^2}/{m^2_\tl},
x_3= {M_3^2}/{m^2_\tq}$) is a loop function whose expression 
can be found {\it e.g.} in Ref.~\cite{lfv-pheno}.
Finally, by using 
eqs.~(\ref{finite},\ref{edmsall}),  we arrive at the main result of our work,  
namely the ratio of the leptonic EDMs can be  predicted only in terms of the 
neutrino parameters: 
\be
\label{edm-lep1}
\frac{d_\mu}{d_e} \!\approx \!\frac{m_\mu}{m_e}
\frac{[\bV (\bm^D)^2_\nu \bV^\dagger]_{22}}
{[\bV (\bm^D)^2_\nu \bV^\dagger]_{11}}
,\quad
\frac{d_\tau}{d_\mu}\!\approx \!\frac{m_\tau}{m_\mu}
\frac{[\bV (\bm^D)^2_\nu \bV^\dagger]_{33}}
{[\bV (\bm^D)^2_\nu \bV^\dagger]_{22}},
\ee
where $d_e \equiv (d_e)_1, d_\mu \equiv (d_e)_2$ {\it etc}, and 
for simplicity we have assumed 
$|\la_1|^2 \ll (\bY^\dagger_T \bY_T)_{ii}$. 
Notice that the presence of extra CPV phases would alter the 
simple form 
of the above ratios (\ref{edm-lep1}) and in general 
the result would be more model dependent. 
Regarding some  numerical insight,  
we can consider three different neutrino mass patterns: the hierarchical
pattern of $m_1 < m_2 \ll m_3$ (HI); the inverted hierarchy of
$m_2>m_1 \gg m_3$ (IH); and the almost degenerate pattern of $m_1 \approx m_2
\approx m_3$ (DG).  For each case, the relative size of the 
entries in eq.~(\ref{edm-lep1}) is given as follows:  
\bea
 {[\bV (\bm^D)^2_\nu \bV^\dagger]_{11}}:
{[\bV (\bm^D)^2_\nu \bV^\dagger]_{22}}:
{[\bV (\bm^D)^2_\nu \bV^\dagger]_{33}}\nonumber \\
 = \cases{  c_{13}^2 s_{12}^2 + \rho s_{13}^2  : \rho c_{13}^2 s_{23}^2
        : \rho c_{13}^2 c_{23}^2  & \mbox{(HI)} \cr
        c_{13}^2 : c_{23}^2 : s_{23}^2 & \mbox{(IH)} \cr
        1 : 1 : 1 & \mbox{(DG}) \cr } \label{YdagY} 
\eea
where $\rho= \!\frac{\Delta m^2_{31}}{\Delta m^2_{21}}\sim 25$ 
 and 
$s_{ij}\, (c_{ij}) \!= \!\sin\theta_{ij} \, (\cos\theta_{ij})$.
Therefore, according to eq.~(\ref{edm-lep1}) and using the present 
best fit neutrino parameters \cite{nuexp} with 
 $s_{13}\ll 0.1$,
we obtain the following leptonic  EDM ratios:
\bea 
 \frac{d_\mu}{d_e} \approx \frac{m_\mu}{m_e}\frac{\rho s^2_{23}
}{s_{12}^2}
 \sim 10^4 , && ~ 
\frac{d_\tau}{d_\mu} \approx \frac{m_\tau}{m_\mu}
\frac{ s^2_{23}}{c^2_{23}}
\sim 17 ,   ~~({\rm HI}) \nonumber \\
\!\!\!\phantom{dd} \frac{d_\mu}{d_e} \approx \phantom{\rho} \frac{m_\mu}{m_e} c^2_{23}
 \sim 10^2 , && ~ 
\frac{d_\tau}{d_\mu} \approx \frac{m_\tau}{m_\mu}
\frac{ s^2_{23}
}{c^2_{23}}
\sim 17 ,   ~~({\rm IH}) \nonumber   \\
\!\!\!\phantom{dd}\frac{d_\mu}{d_e} \approx \frac{m_\mu}{m_e} 
 \sim 2\times 10^2 , && ~
\!\!\!\! \frac{d_\tau}{d_\mu} \approx \frac{m_\tau}{m_\mu}
\sim 17 ,   ~~({ \rm DG}) .  \label{edmHI}
\eea

We are now tempted to  give an order-of-magnitude estimate
of $d_e$ to show that sizeable
values can be attained:
\be
\frac{d_e}{e} \sim  10^{-29} 
\left(\frac{M_T}{10^{11}~{\rm GeV}}\cdot
\frac{10^{-4}}{\lambda_2}\right)^2
\left ( \frac{200~{\rm GeV}}{\tm}\right)^{2}~
{\rm cm}
\ee
where  we have taken a common SUSY mass scale, 
$M_1 =  m_\tl ={\rm Im}(B_T)= \tm $ and the pattern HI.
This shows that the electron and muon EDMs could be within
the future experiment reach (see Table~1). 
We also notice from eq.~(\ref{finite}) 
that lepton and quark EDMs are definitely correlated 
in this scenario. However,  such a correlation is also sensitive 
to the ratio $M_T/\lambda_2$ and to  
other mass parameters, such as the gaugino,  squark and slepton 
masses, and so can only be established in a specific SUSY breaking 
framework. 

Another interesting prediction  
regards 
the relative size of LFV among different flavours \cite{ar}.  
For instance, the ratio of the LFV entries of the left-handed 
slepton mass matrix is: 
\be
\label{predi}
 \frac{ (\bm^{2 }_{\tilde{L}})_{\tau \mu}}
  {(\bm^{2 }_{\tilde{L}})_{\mu e} } \approx 
\frac{ (\bY^{\dagger}_T \bY_T)_{23} }
  { (\bY^{\dagger}_T \bY_T)_{12} } \approx 
\rho
\frac{\sin 2\theta_{23}}{\sin 2\theta_{12}
\cos\theta_{23}} \sim 40
\ee
which holds for  $s_{13} \ll \rho^{-1} c_{12} s_{12} \sim 0.02 $.
This implies that also the branching ratios 
$B(\ell_i \to \ell_j  \gamma)$
can be related in terms of only the low-energy 
neutrino parameters and  we find
\be\label{brs}
{\rm B}(\mu \to e \gamma) : {\rm B}(\tau \to e \gamma) :
{\rm B}(\tau \to \mu \gamma) \sim 1 : 10^{-1} : 300 .
\ee
This result does not depend
on the detail of the model, such as either $M_T$
or the SUSY spectrum \cite{ar}. 
On the contrary, the individual branching ratios in (\ref{brs}) also 
depend on quantities such as $\mu, \tan\beta$ and soft SUSY parameters, 
which are not of direct concern in the present discussion of the EDMs.

\begin{table}
\begin{ruledtabular}
\begin{tabular}{ccc}
EDM & Present limits & Future limits \\[0.2pt]
\hline
$d_e$ &  $7\times 10^{-28}$ \cite{pdg} &  $10^{-32}$ \cite{de-f}  \\
$d_\mu$ & $3.7\times 10^{-19}$ \cite{pdg}& 
$10^{-24} - 5\times 10^{-26}$ \cite{dmu-f} \\
${\rm Re}(d_\tau)$  & $4.5\times 10^{-17}$ \cite{pdg} & $10^{-17} - 10^{-18}$ 
\\ \hline
$d_n$&  $6\times 10^{-26}$ \cite{pdg} & ?? \\
\end{tabular}
\end{ruledtabular}
\caption{Present bounds and future sensitivity (in {\rm e$\cdot$cm} units) 
on lepton and neutron EDMs.
}
\label{tb3}
\end{table}

The presence of triplet states with mass smaller than 
the grand unification scale $M_G$
precludes the gauge-coupling unification.
This can be recovered, for instance,
by completing a GUT representation
where $T, \bar{T}$  fit. 
 In such a case care should be taken in evaluating the 
radiative effects as the additional components of the full GUT multiplet  
where triplets reside would also contribute to LFV as well as CPV processes.  
For an explicit example, see, for instance,
Ref.\cite{ar}.

We have focused on the generation of lepton and quark EDMs in 
the triplet-seesaw scenario by assuming that only the soft
parameter  $B_T$  has a nonzero phase. 
Though we believe that realistic superymmetry-breaking mechanisms 
can realize such a scenario, we have not addressed this 
interesting issue in this Letter.   
We find that the inter-family ratios 
$d_\mu/d_e$ and $d_\tau/d_\mu$  are determined only by 
the  low-energy neutrino parameters and so are independent of 
the details of the model (such as the SUSY spectrum or the ratio 
$M_T/\lambda_2$).
The leptonic EDMs are also correlated with the quark EDMs.
This correlation is somewhat more model dependent, 
since it may depend {\it e.g.} on the SUSY breaking scenario 
and therefore it would deserve a more detailed analysis
which is beyond the scope of this work.
This model dependence is also present   
in the correlation  among the fermion EDMs and 
LFV processes, such as $\mu \to e \ga, \tau \to \mu\ga$.
Nevertheless, the  EDM relations (\ref{edm-lep1})  
and the radiative-decay  BR ratios  (\ref{brs}) 
are unique signature to test the triplet-seesaw scenario 
in the upcoming experiments.
Finally, we recall another role played by the 
term $B_T$ in the context of  `soft' leptogenesis 
\cite{tri-lepto}. 
Namely, leptogenesis can be realized with only a single 
pair of triplets $T,\bT$, via the $L$ violating decays 
of the latter, thanks to the resonant enhancement 
induced by the mass splitting $B_T M_T$ between the triplet mass eigenstates. 
Moreover, in this picture  CP asymmetries arise from nonzero   
relative phases among the superpotential $W_T$ and related 
soft trilinear couplings \footnote{The complex 
$B_T$-term also induces finite corrections to the other soft-trilinear terms 
related to the couplings $\bY_T, \la_1, \la_2$ in $W_T$ of eq.~(\ref{Wt}). 
In this way, an unremovable CP phase emerges among those related couplings.}.

\medskip

\mysection{Acknowledgements:} E.~J.~C was supported by the Grant, 
KRF-2002-070-C00022. The work of A.~R.~ has been partially 
supported by the EU HPRN-CT-2000-00148 (Across the Energy Frontier) 
and HPRN-CT-2000-00149 
(Collider Physics) contracts. S.~K.~V. acknowledges support 
from the Italian MIUR 
under the program `PRIN:Astroparticle Physics´ 
2002 and Indo-French Centre 
for Promotion of Advanced Research (CEFIPRA) 
project No:  2904-2 `Brane World´ Phenomenology.  
He is also partially supported by
INTAS grant, 03-51-6346, CNRS PICS \# 2530,
RTN contracts MRTN-CT-2004-005104 and
MRTN-CT-2004-503369 and by a European Union Excellence Grant,
MEXT-CT-2003-509661.

\end{document}